\documentstyle[11pt,IAUS215,twoside,epsf]{article}

\def\msun{\,\mathrm{M}_\odot}
\def\mso{\,\mathrm{M}_\odot}
\def\msoy{\, \mso~{\rm yr}^{-1}}
\def\simle{\mathrel{\hbox{\rlap{\hbox{\lower4pt\hbox{$\sim$}}}\hbox{$<$}}}}
\def\simgr{\mathrel{\hbox{\rlap{\hbox{\lower4pt\hbox{$\sim$}}}\hbox{$>$}}}}
\def\msoy{\, \mso~{\rm yr}^{-1}}

\def\edcomment#1{\iffalse\marginpar{\raggedright\sl#1\/}\else\relax\fi}
\reversemarginpar

\begin{document}
\vspace*{1cm}
\title{Binary evolution models with rotation}
\author{N. Langer, S.-C. Yoon, J. Petrovic}
\affil{Astronomical Institute, P.O. Box 80000, NL-3508 TA Utrecht, 
  The Netherlands}
\author{A. Heger}
\affil{Department of Astronomy and Astrophysics, Enrico Fermi
  Institute, The University of Chicago, 5640 S. Ellis Avenue, Chicago
  IL 60637, USA}

\begin{abstract}
We discuss the first available binary evolution models which include
up-to-date rotational physics for both components, as well as angular momentum
accretion and spin-orbit coupling. These models allow a self-consistent
computation of the mass transfer efficiency during Roche-lobe overflow
phases, and a determination of the transition from quasi-conservative to
non-conservative evolution. Applications to massive binary systems lead to
predictions for the spin rates of compact objects in binaries, and for
the occurrence of gamma-ray bursts from collapsars in binaries. 
Rotational effects in
accreting white dwarfs are found to stabilise the shell burning and
decrease the carbon abundance in progenitor models for Chandrasekhar-mass  
Type~Ia supernovae, and to potentially avoid a detonation of the white
dwarf within the sub-Chandrasekhar mass scenario.
\end{abstract}

\section{Introduction}
The evolution of a single star can be strongly influenced by
its rotation (e.g., Heger \& Langer 2000; Meynet \& Maeder
2000), and evolutionary models of rotating stars are 
now available for many masses and metallicities.
While the treatment of the rotational processes in these models
is not yet in a final stage (e.g., magnetic dynamo processes are just 
about to be included; Heger et al., this volume), they provide
first ideas of what rotation can really do to a star.

Effects of rotation, as important they are in single stars,
can be much stronger in the components of close binary systems: 
Estimates of the angular momentum gain of the accreting star
in mass transferring binaries show that critical rotation
may be reached quickly (Packet 1981; Langer et al. 2000).
Therefore, we need binary evolution models
which include a detailed treatment of rotation in the stellar interior, 
as in recent single star models. However, in binaries, tidal
processes as well as angular momentum accretion need to be considered
at the same time. Some first such models are now available and
are discussed below.

These models provide evidence for rotational processes in binaries
being essential for some of the most exciting cosmic
phenomena, which may occur exclusively in binaries: Type~Ia supernovae,
the main producers of iron and cosmic yardstick to measure the
accelerated expansion of the universe. Gamma-ray bursts from collapsars
--- which current stellar models with rotation
preclude to occur in single stars (Heger et al., this volume;
Woosley \& Heger, this volume)
--- may provide the most powerful explosions
in the universe and trace star formation to its edge. And black holes,
even though also formed by single stars, need a companion to become
``visible''.

\section{How much matter can stars accrete from a binary companion?}

As mentioned above, non-magnetic accretion, i.e. accretion via a viscous disk or
via ballistic impact, transports angular momentum and can lead to a strong
spin-up of the mass gaining star. For disk accretion, it appears plausible that
the specific angular momentum of the accreted matter corresponds to
Kepler-rotation at the stellar equator; this leads to a spin-up of 
the whole star to critical rotation when its initial mass is increased
by about 20\% (e.g., Packet 1981). Can accretion continue beyond this?

Theoretically, this appears possible, as viscous processes may 
transport angular momentum 
outwards through the star, the boundary layer, and the accretion
disc (e.g., Paczynski 1991). 
However, as the star may be strongly rotating, its wind mass
loss may be dramatically increased (Langer 1997, 1998), which may render
the mass transfer process inefficient.

Observations of massive post-mass transfer binary systems constrain this
effect.
Table~1 lists parameters of four different kinds of massive close
binary systems which give opposite answers. The two O~stars in the
Case~A binary (mass transfer starts while both stars undergo core
hydrogen burning)
V$\,$729~Cyg have a mass ratio of 3.5 but the same
spectral type and visual flux. Clearly, an initial mass ratio close
to~1 is required to get close to the observed current mass ratio.
However, as during Case~A the primary star (as we designate the
initially more massive star in a binary) loses just about half of
its mass, a mass ratio of at most~2 could be produced were the
secondary (the initially less massive star in a binary) 
not allowed to accrete. Another system showing strong evidence
for accretion is the massive X-ray binary Wray~977; it would require
that stars of $\ga 40\mso$ form neutron stars to explain this system
without accretion (Wellstein \& Langer 1999).

\begin{table}[t]
\caption{Clues on accretions efficiencies from observed binaries}
\begin{tabular}{c c c c c}
\tableline
object & sp. types & orb. period & masses/ratio & accretion? \\ 
\tableline
V729~Cyg$^1$ & O7+O7$^2$ & 6.6$\,$d & $q=3.5$ & YES \\
Wray 977$^3$ & BI+NS & 44$\,$d  & $40\msun + 1.4\msun$ & YES \\
3 systems$^4$& WNE+O & $\sim 8\,$d& $q\simeq 0.5$ & NO$^5$ \\
4U~1700-37$^6$& O6I+NS/BH& 3.4$\,$d & $58\msun + 2.4\msun$ & NO$^5$ \\
\tableline
\tableline
\end{tabular}\\
$^1$  Bohannan \& Conti (1976) \\
$^2$  both components have the same visual magnitude \\
$^3$  Kaper et al. (1995), Wellstein \& Langer (1999) \\
$^4$  Petrovic \& Langer (2002) \\
$^5$  meaning: 10\% or less \\
$^6$  Clark et al. (2002) \\
\end{table}
\index{V729~Cyg}\index{Wray 977}\index{4U~1700-37}

Several Galactic short period WNE+O binaries, on the other hand, can
not be understood had the O~star accreted substantial amounts 
from the WNE progenitor
(Petrovic \& Langer 2002). While those might have formed through
common envelope evolution --- for which little accretion is expected --- the key
X-ray binary 4U$\,$1700-37 has such a short period that a major
accretion phase can be excluded. However, as Case~C evolution
(mass transfer starts after core helium burning of the primary
star) would 
lead to a compact object much more massive than 2.4$\mso$, Case~B
evolution
(mass transfer starts just after core hydrogen burning of the primary
star)
is most likely here (Clark et al. 2002).

We conclude from these observed binary systems 
that in some massive close (Case~A or~B) 
binaries, the mass
transfer process is nearly conservative, while in others it is strongly
non-conservative. In the following, we show that including rotational
physics into binary evolution calculations allows to recover both features
within the same physical model.

\section{Binary models with rotating components}

We have constructed binary evolution models using the code of Wellstein
et al. (2001), but including the physics of rotation as in the single
star models of Heger et al. (2000) for both components. 
In addition, spin-orbit coupling according to Zahn (1977) has been
added, and rotationally enhanced winds are implemented as in Langer
(1998). The specific angular momentum of the accreted matter is assumed
to be that of Kepler rotation at the stellar equator in the case of
disk accretion, and determined by integrating the equation of motion
of a test particle in the Roche potential in case the accretion stream
impacts directly on the secondary star (Wellstein 2001).

\begin{figure}[th!]
\centering
\epsfxsize=0.88\hsize
\centerline{\epsffile{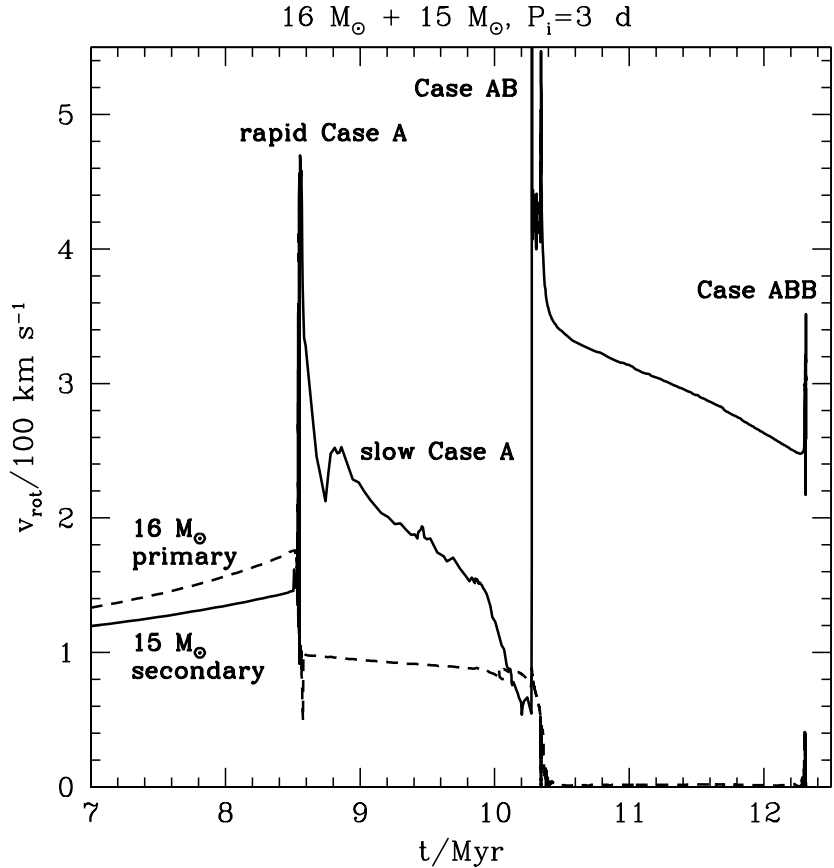}}
\caption{Equatorial rotation velocity for primary
(dashed line) and se\-condary (solid line) component
of a $16\mso +15\mso$ system with an initial orbital period of
3~d as function of time, starting at an age of 7~Myr, i.e., well before
the onset of mass transfer, until the end of Case~ABB mass transfer,
which corresponds to the time of the supernova explosion of the primary.
The four different mass transfer phases which occur in this system
are indicated; except for
the slow Case~A mass transfer they occur on the thermal time scale
of the primary star (see also Fig.~2).
}
\end{figure}

Fig.~1 shows the evolution of the equatorial rotation velocity in a
system starting out with a $16\mso$ and a $15\mso$ star in a 3~day
orbit. The initial rotational velocity of both stars is unimportant
since they evolve quickly into rotation which is synchronous with the orbital
revolution, due to spin-orbit coupling. 
Each of the three thermal time scale mass transfer phases through
which this system evolves (rapid Case~A, Case~AB, and Case~ABB; 
see Wellstein et al. 2001) leads to a strong spin-up of the secondary
star and an equally drastic spin-down of the primary (see Langer 1998,
for the purely mechanical spin-down effect). 

\begin{figure}[t]
\centering
\epsfxsize=0.70\hsize
\centerline{\epsffile{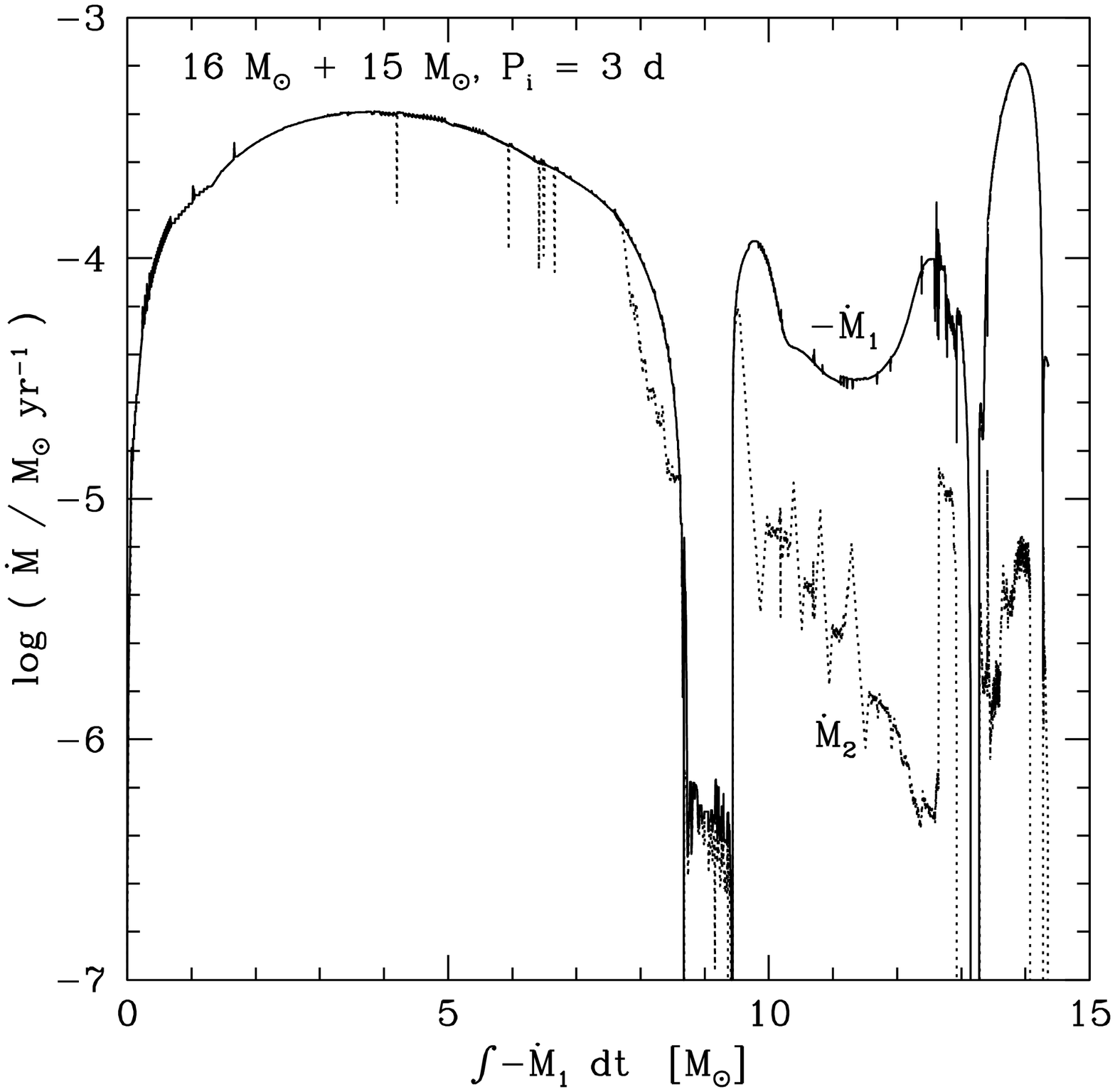}}
\caption{For the $16\mso +15\mso$ system with an initial orbital period of
3~days: mass transfer rate ($-\dot M_{\rm 1}$; full drawn line) 
and mass accretion
rate of the secondary ($\dot M_{\rm 2}$; dotted line) as function of the 
total amount of mass which has already been transferred.
Four discrete mass transfer phases can be distinguished:
rapid Case~A ($-\dot M_{\rm 1}\ga 10^{-4}\msoy$, 
$\Delta M_{\rm 1}\simeq 8.7\msun$),
slow Case~A ($-\dot M_{\rm 1}\ga 10^{-7}\msoy$, 
$\Delta M_{\rm 1}\simeq 0.8\msun$),
Case~AB ($-\dot M_{\rm 1}\ga 10^{-5}\msoy$, 
$\Delta M_{\rm 1}\simeq 3.8\msun$), and
Case~ABB ($-\dot M_{\rm 1}\ga 10^{-4}\msoy$,
$\Delta M_{\rm 1}\simeq 1.2\msun$).
}
\end{figure}

The accretion induced spin-up can bring the secondary close to critical
rotation and thus strong rotationally enhanced mass loss sets in. The
difference between the mass overflow rate $-\dot M_{\rm 1}$ and the
wind mass loss rate of the secondary is the net accretion rate 
$\dot M_{\rm 2}$ of the secondary star. Fig.~2 shows that during much
of the rapid Case~A mass transfer --- during which the accreting star
increases its mass from 15$\mso$ to $\sim 23\mso$ ---, 
it is $-\dot M_{\rm 1} = \dot M_{\rm
2}$. This is possible due to two factors: 1)~During direct impact
accretion, the specific angular momentum of the accreted material is
only a fraction of the respective Kepler angular momentum. 2)~As the
secondary star fills a significant fraction of its Roche volume, tidal
coupling removes part of the accreted angular momentum and feeds it
into the orbit. 

As the orbit widens significantly during the later
evolution, the specific angular momentum of the accreted matter
increases and tidal forces weaken. Thus, while the accretion efficiency
is close to~1 during Case~A, it is less than~0.1 later, resulting in a
time average over the whole evolution of~0.67. For systems which
start out with a wider orbit, we find a low accretion efficiency
throughout (Fig.~3). For $16\mso +15\mso$ systems, we find a critical
initial period of $\sim 8\,$d beyond which the secondary accretes only
little. 

\begin{figure}[t]
\centering
\epsfxsize=0.92\hsize
\centerline{\epsffile{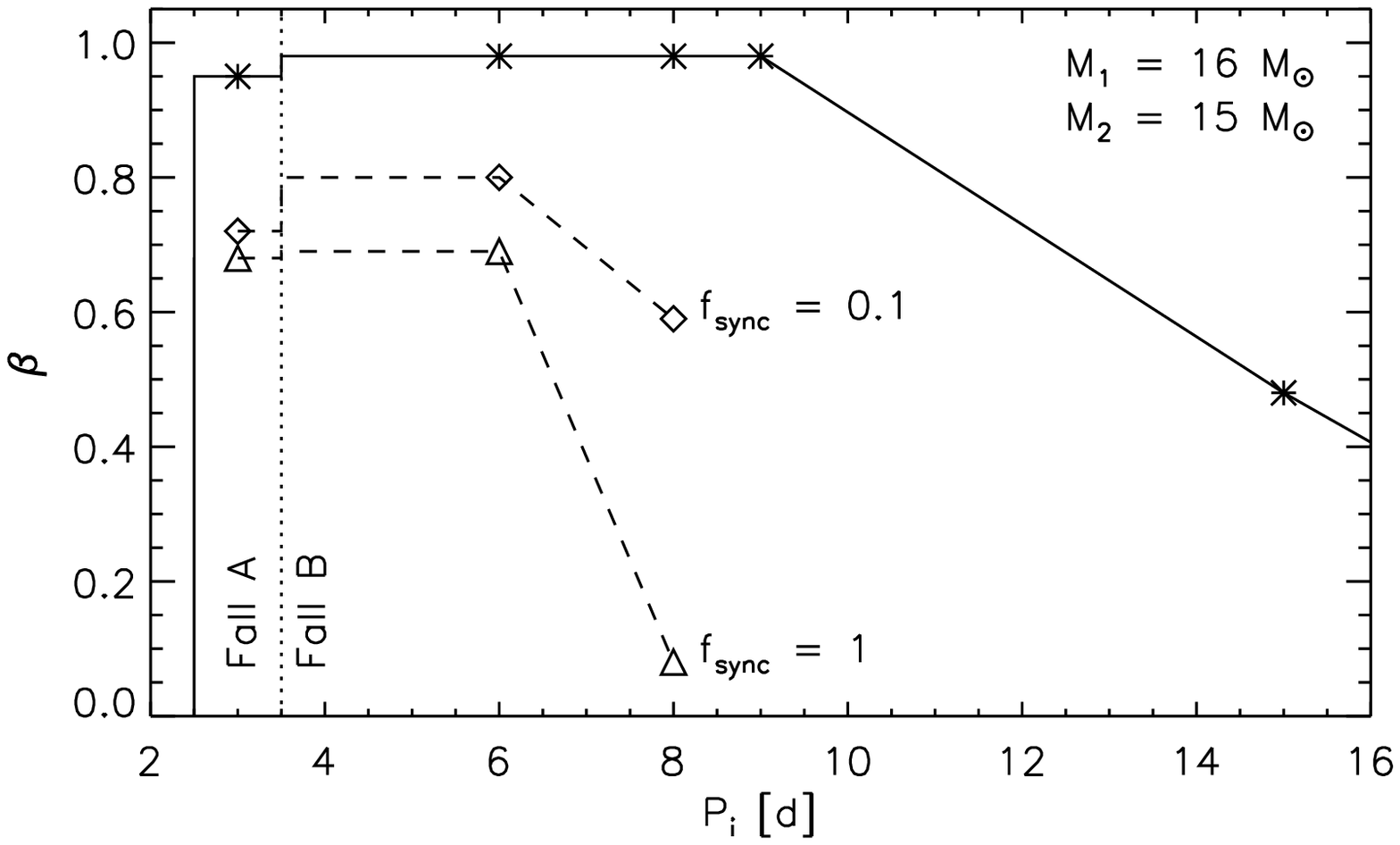}}
\caption{Time averaged mass accretion efficiency $\beta$ for various
binary evolution sequences (symbols) and different physical assumptions,
as function of the initial orbital period for $16\mso +15\mso$ systems.
Triangles mark systems which have been calculated using rotational
physics and the tidal model as proposed by Zahn (1977), which in systems
marked by squares has been assumed to be one order of magnitude more
efficient. Star symbols designate models by Wellstein et al. (2001)
which have been computed without rotational physics and assumed to
evolve conservatively unless a contact situation occurs. The dotted
vertical line separates Case~A systems (to the left) from Case~B
systems.  }
\end{figure}

These models imply that, despite the angular momentum problem 
for the accretion star (Packet 1981), quasi-conservative evolution of
massive close binaries is possible. However, already in early Case~B
systems, the accretion efficiency may be strongly reduced compared to
binary models without rotation.

\section{Black hole formation in binaries}

While single stars more massive than $\sim 25\mso$ are supposed to form
black holes (Fryer 2002), it is a challenge to binary
evolution modellers to find a way to make black hole binaries ---
binary systems consisting of a non-degenerate star and a black hole.
The reason is that the initially more massive star, which is supposed
to end up as black hole, undergoes tremendous mass loss (Fig.~2,
and see Wellstein \& Langer 1999). 

Brown, Lee \& Bethe (1999) suggested the Case~C channel to form
the Galactic low mass BH binaries --- a low mass star
in close orbit with a black hole of typically 6$\mso$ (Orosz 
2002), i.e., starting with a wide system consisting of a (say)
25$\mso$ and a $5\mso$ star. This suggestion is strongly supported by
the massive post-Case~B X-ray binary 4U$\,$1700-37 (Table~1): 
it shows that in Case~B
systems, even a 60$\mso$ primary forms a compact object of only
2.4$\mso$ (Clark et al. 2002). 

The problem of how wide the Case~C
channel is for such massive stars is difficult. Current single star
models suggest no significant radius increase late during or after core
helium burning at solar metallicity. On the other hand, 
Smartt's (2002) abundance analysis of Galactic supergiants 
suggests that they do not undergo ``blue loop'' evolution --- which
contradicts the current models. The problem is less severe at low
metallicity, where many models predict a gradual radius increase during
core helium burning.

It is even harder to produce Cyg~X-1 type BH binaries, with black hole
masses as high as 14$\mso$. Brown et al. (2001) argue that this
requires Case~C evolution and an upturn of the initial-final mass
relation for single stars above $\sim 60\mso$, as predicted by
Langer (1987).

\section{Spins of compact objects and gamma-ray bursts}

The inclusion of rotationally enhanced magnetic fields in the evolution
of massive single stars (Heger et al., this volume) has improved the
agreement between observed ($15 ... 150\,$ms) and predicted ($\simgr 5\,$ms) 
spin periods of young neutron stars --- with the consequence that
obtaining gamma-ray bursts from black hole formation in single stars
(collapsars; Woosley \& Heger, this volume) seems difficult at
present.

From Sections~3 and~4 above, we conclude that the initially more
massive stars in massive close binaries are even less likely to produce a
gamma-ray burst. First of all, they lose so much mass that even stars
with a very large initial mass may not even form a black hole but rather
a neutron star (see also Wellstein \& Langer 1999). And secondly, Figure~1
shows how drastic these stars are spun down as a consequence of their
heavy mass loss. The 16$\mso$ star in the computed binary system is
expected to produce a neutron star with an initial spin period of more
than one second!

The only way to avoid both drawbacks is to employ Case~C evolution,
which leads to a core evolution as in single stars. However, even then
the CO-core of the star needs to be spun-up significantly to produce
a collapsar and a gamma-ray burst --- a possibility suggested by Brown et
al. (1999, 2000) in the context of common envelope evolution and
spiral-in. No detailed models for this scenario exist at present.

The initially less massive star in a massive binary, on the other hand,
accretes large amounts of angular momentum and will thus acquire a larger
core spin than a corresponding single star (see Fig.~1). 
It is thus conceivable that accretion stars are the progenitors of
asymmetric supernova explosions and rapidly spinning compact objects.
Those sufficiently massive to transform into a Wolf-Rayet star during core
helium burning, or those which lose their envelope in a reverse Case~C
mass transfer,
may possess all required ingredients to produce a
gamma-ray burst within the collapsar model. 

\section{Progenitors of Type~Ia supernovae}

Supernovae of Type~Ia are supposed to occur in close binary systems where
the accreting component is a CO-white dwarf. The currently favoured
scenario assumes that a white dwarf of initially less than $1\mso$
grows in mass due to quiescent nuclear shell burning of hydrogen and/or
helium, which is accreted from a non-degenerate companion.
In order to explode, it must reach the Chandrasekhar-mass, i.e., it needs
to accrete at least $0.4\mso$. While single white dwarfs are slow
rotators (Kawaler, this volume), the idea that Type~Ia supernova
progenitors are significantly spun-up due to accretion appears to be
supported by recent measurements of the rotational velocity of the white
dwarf in cataclysmic variables (Starrfield, this volume).

In order for the degenerate CO-core to grow in mass, accreting white
dwarfs need to possess a helium burning shell source. However, like in
AGB stars, which have a degenerate CO-core in their center, the helium
shell source in accreting white dwarfs tends to turn thermally unstable,
causing large oscillations in nuclear energy production, as well as in
the white dwarf radius (Fig.~4). It is unclear whether the binary system can
survive these oscillations intact.
\begin{figure}[t]
\epsfxsize=0.88\hsize
\centerline{\epsffile{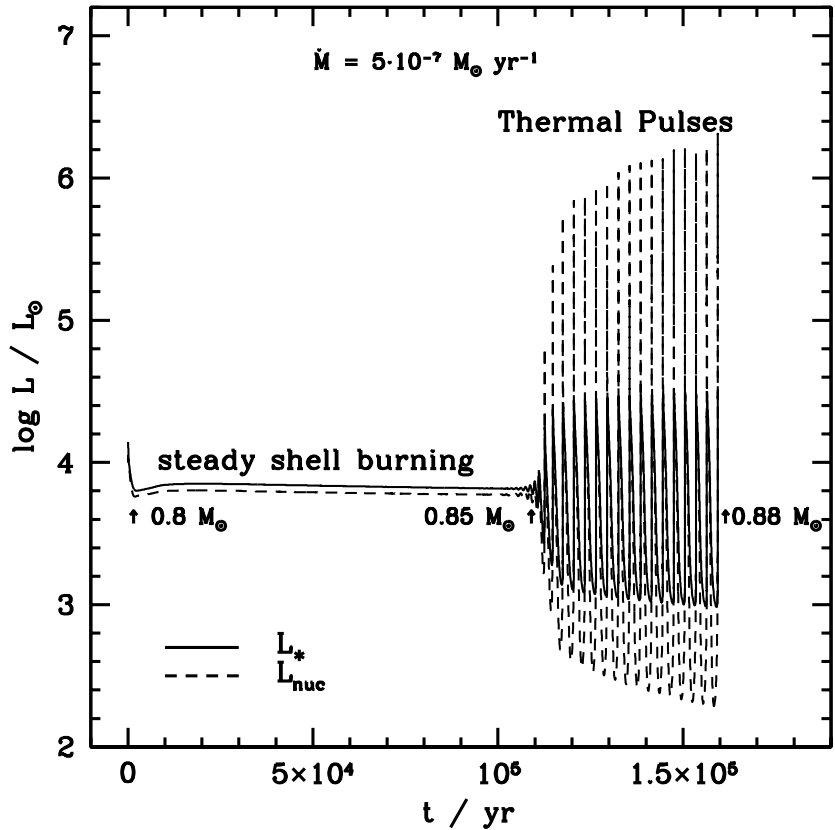}}
\caption{Stellar and nuclear luminosity as function of time for a
C/O-white dwarf model starting at $0.8\mso$, which accretes helium
at a constant rate of $5\times 10^{-7}\msoy$ (Scheithauer 2000).  }
\end{figure}
\begin{figure}[t]
\epsfxsize=0.8\hsize
\centerline{\epsffile{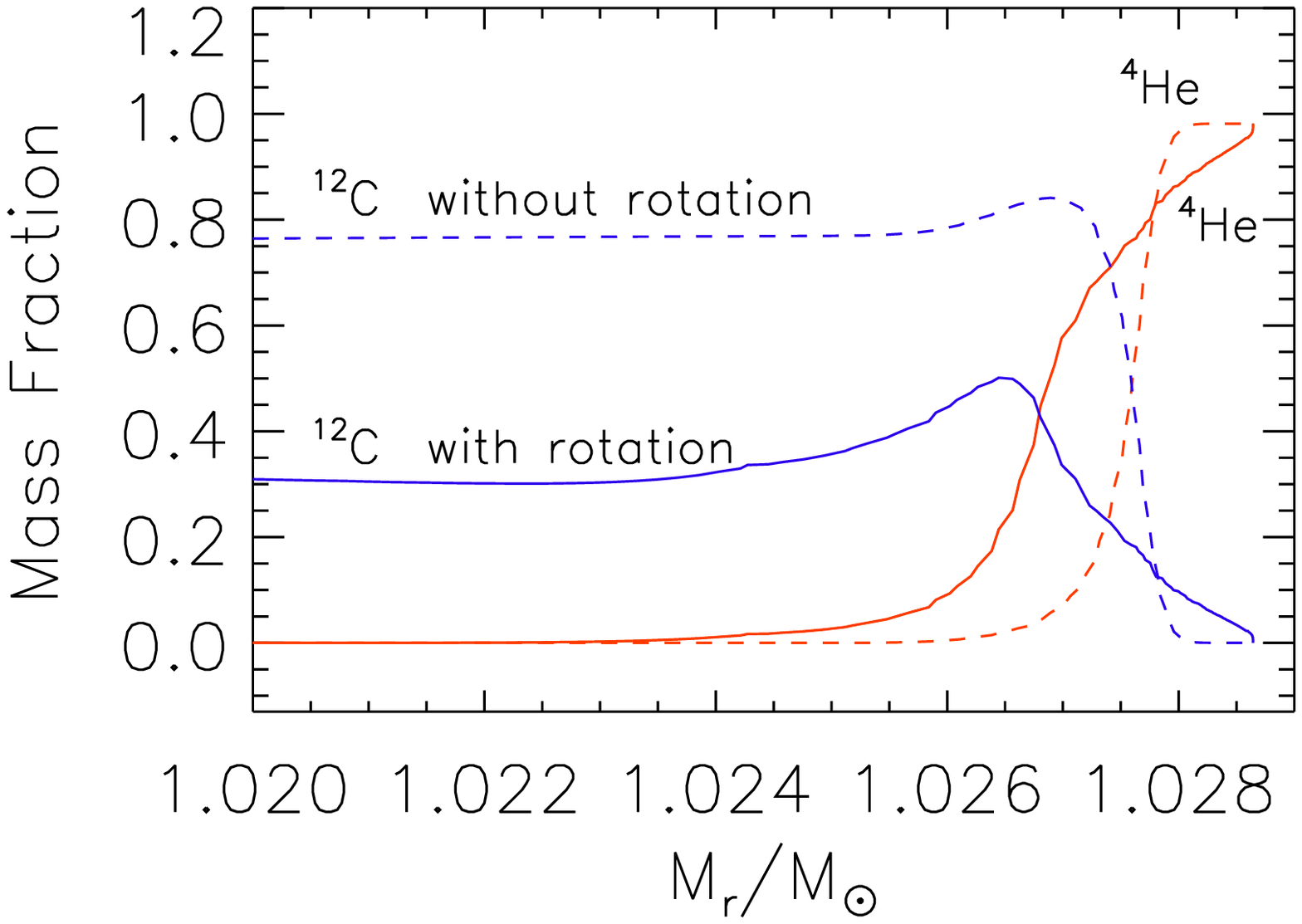}}
\caption{Helium and carbon mass fraction as
function of the mass coordinate in a CO white dwarf which starts
at $1\mso$ and accretes helium-rich matter at a rate of $10^{-6}\msoy$,
at a white dwarf mass of  $1.0286\mso$, computed with and without
rotation (Yoon \& Langer 2002). Note that the carbon mass
fraction $X_{\rm C}$ behind the shell source ($M_{\rm r} < 1.025\mso$)
is lower in the rotating
model ($X_{\rm C}\simeq 0.3$) than in the non-rotating one
($X_{\rm C}\simeq 0.8$), corresponding to an
oxygen mass fraction of 0.7 rather than 0.2.}
\end{figure}
However, Yoon et al. (2003) found that the consideration that angular
momentum is accreted along with the matter by the white dwarf results in
different shell source properties. The spin-up of the outer part of the
white dwarf leads to a lower effective gravity and to a more relaxed
helium shell source. Additionally, differential rotation leads to
shear mixing at the location of the shell source, increasing its
geometrical thickness and often turning it stable. This effect increases
the likelihood of a given system to reach the Chandrasekhar-mass at all.

However, not only does the thickness of the helium shell source
increase, but the chemical profiles through it become shallower (Fig.~5).
As thus most of the helium burning occurs at a lower helium
concentration, oxygen is produced at a much higher abundance than in
corresponding non-rotating models, and carbon is produced
correspondingly less. 
As the combustion of oxygen yields less energy than that of carbon,
rotational effects influence the brightness of Type~Ia
supernovae. As the rotation rate depends on the amount of accreted
material, and the average amount of accreted matter increases with the
initial metallicity of the binary system (Langer et al. 2000), a drop of
the brightness of Type~Ia supernovae with higher metallicity  ---
or an increase with redshift --- might occur.

Finally, rotation also affects the evolution
of white dwarfs in the so called sub-Chandrasekhar mass scenario, as
outlined by Yoon \& Langer (2003; also this volume).



\end{document}